\documentclass{emulateapj}

\newcommand{\Kepler}{{\it Kepler}}
\newcommand{\Spitzer}{{\it Spitzer}}
\newcommand{\Hipparcos}{{\it Hipparcos}}
\newcommand{\Tycho}{{\it Tycho}}
\newcommand{\emcee}{{\it emcee}}

\newcommand{\ms}{m~s$^{-1}$}
\newcommand{\kms}{km~s$^{-1}$}

\newcommand{\thisstar}{HIP\,116454}
\newcommand{\thisplanet}{HIP\,116454\,b}

\newcommand{\mearth}{M$_\oplus$}
\newcommand{\rearth}{R$_\oplus$}
\newcommand{\msun}{M$_\odot$}
\newcommand{\rsun}{R$_\odot$}
\newcommand{\feh}{[Fe/H]}
\newcommand{\fe}{-0.16}
\newcommand{\ufe}{0.08}
\newcommand{\loggst}{4.590} 
\newcommand{\uloggst}{0.026}
\newcommand{\teff}{5089}
\newcommand{\uteff}{50}

\newcommand{\meh}{[M/H]}
\newcommand{\water}{H$_{2}$O}
\newcommand{\mgsi}{MgSiO$_3$}
\newcommand{\mst}{0.775}
\newcommand{\umst}{0.027}
\newcommand{\rst}{0.716}
\newcommand{\urst}{0.024}

\newcommand{\rhost}{2.11}
\newcommand{\urhost}{0.23}

\newcommand{\distance}{55.2}
\newcommand{\udistance}{5.4}

\newcommand{\rprst}{0.0311}
\newcommand{\urprst}{0.0017}
\newcommand{\depth}{0.000967}
\newcommand{\udepth}{0.000109}
\newcommand{\arst}{27.22}
\newcommand{\uarst}{1.14}
\newcommand{\incl}{88.43}
\newcommand{\uincl}{0.40}
\newcommand{\imp}{0.65}
\newcommand{\uimp}{0.17}
\newcommand{\ecc}{0.205}
\newcommand{\uecc}{0.072}
\newcommand{\omegap}{-59.1}
\newcommand{\uomegap}{16.7}
\newcommand{\semi}{4.41}
\newcommand{\usemi}{0.50}
\newcommand{\mpl}{11.82}
\newcommand{\umpl}{1.33}
\newcommand{\rpl}{2.53}
\newcommand{\urpl}{0.18}
\newcommand{\perpl}{9.1205}
\newcommand{\uperpl}{0.0005}
\newcommand{\rhopl}{4.17}
\newcommand{\urhopl}{1.08}
\newcommand{\teq}{690}
\newcommand{\uteq}{14}
\newcommand{\loggpl}{3.26}
\newcommand{\uloggpl}{0.08}
\newcommand{\ttransit}{2456907.89}
\newcommand{\uttransit}{0.03}

\usepackage[bookmarks=false]{hyperref}

\shorttitle{The First K2 Planet}
\shortauthors{Vanderburg et al.}

\begin{document}


\title{Characterizing K2 Planet Discoveries:\\ A super-Earth transiting the bright K-dwarf \thisstar}


\author{
Andrew Vanderburg\altaffilmark{$\star$,1,2}, 
Benjamin T. Montet\altaffilmark{1,2,3}, 
John Asher Johnson\altaffilmark{1,4}, 
Lars A. Buchhave \altaffilmark{1}, 
Li Zeng\altaffilmark{1}, 
Francesco Pepe\altaffilmark{8},
Andrew Collier Cameron\altaffilmark{7},
David W. Latham\altaffilmark{1}, 
Emilio Molinari\altaffilmark{9,10},
St\'{e}phane Udry\altaffilmark{8},
Christophe Lovis\altaffilmark{8},
Jaymie M. Matthews\altaffilmark{16},
Chris Cameron\altaffilmark{20},
Nicholas Law\altaffilmark{5}, 
Brendan P. Bowler\altaffilmark{3,19},
Ruth Angus\altaffilmark{1,15},
Christoph Baranec\altaffilmark{6}, 
Allyson Bieryla\altaffilmark{1},
Walter Boschin\altaffilmark{9},
David Charbonneau\altaffilmark{1},
Rosario Cosentino\altaffilmark{9},
Xavier Dumusque\altaffilmark{1},
Pedro Figueira\altaffilmark{28,29},
David B. Guenther\altaffilmark{21},
Avet Harutyunyan\altaffilmark{9},
Coel Hellier\altaffilmark{18},
Rainer Kuschnig\altaffilmark{22},
Mercedes Lopez-Morales\altaffilmark{1},
Michel Mayor\altaffilmark{8},
Giusi Micela\altaffilmark{12},
Anthony F. J. Moffat\altaffilmark{23,24},
Marco Pedani\altaffilmark{9},
David F. Phillips\altaffilmark{1},
Giampaolo Piotto\altaffilmark{9,10},
Don Pollacco\altaffilmark{17},
Didier Queloz\altaffilmark{11},
Ken Rice\altaffilmark{13},
Reed Riddle\altaffilmark{3}, 
Jason F. Rowe\altaffilmark{25,26},
Slavek M. Rucinski\altaffilmark{27},
Dimitar Sasselov\altaffilmark{1},
Damien S\'{e}gransan\altaffilmark{8},
Alessandro Sozzetti\altaffilmark{30},
Andrew Szentgyorgyi\altaffilmark{1},
Chris Watson\altaffilmark{14},
Werner W. Weiss\altaffilmark{22}
}

\altaffiltext{$\star$}{avanderburg@cfa.harvard.edu}
\altaffiltext{1}{Harvard-Smithsonian Center for Astrophysics, Cambridge, MA 02138, USA}
\altaffiltext{2}{NSF Graduate Research Fellow}
\altaffiltext{3}{California Institute of Technology, Pasadena, CA, 91125, USA}
\altaffiltext{4}{David \& Lucile Packard Fellow}
\altaffiltext{5}{University of North Carolina at Chapel Hill, Chapel Hill, NC, 27599, USA}
\altaffiltext{6}{University of Hawai`i at M\={a}noa, Hilo, HI 96720, USA}
\altaffiltext{7}{SUPA, School of Physics and Astronomy, University of St Andrews, North Haugh, St Andrews, Fife KY16 9SS, UK}
\altaffiltext{8}{Observatoire Astronomique de l'Universit\'e de Gen\`eve, 51 chemin des Maillettes, 1290 Versoix, Switzerland. }
\altaffiltext{9}{INAF - Fundaci\'on Galileo Galilei, Rambla Jos\'e Ana Fern\'andez P\'erez, 7, 38712 Bre\~na Baja, Spain}
\altaffiltext{10}{INAF - IASF Milano, via Bassini 15, 20133, Milano, Italy}
\altaffiltext{11}{Cavendish Laboratory, J J Thomson Avenue, Cambridge CB3 0HE, UK}
\altaffiltext{12}{INAF - Osservatorio Astronomico di Palermo, Piazza del Parlamento 1, 90124 Palermo, Italy}
\altaffiltext{13}{SUPA, Institute for Astronomy, Royal Observatory, University of Edinburgh, Blackford Hill, Edinburgh EH93HJ, UK}
\altaffiltext{14}{Astrophysics Research Centre, School of Mathematics and Physics, Queen’s University Belfast, Belfast BT7 1NN, UK}
\altaffiltext{15}{University of Oxford, Oxford, UK}
\altaffiltext{16}{University of British Columbia, Vancouver, BC, V6T1Z1, Canada}
\altaffiltext{17}{Department of Physics, University of Warwick, Gibbet Hill Road, Coventry CV4 7AL, UK}
\altaffiltext{18}{Astrophysics Group, Keele University, Staffordshire ST5 5BG, UK}
\altaffiltext{19}{Caltech Joint Center for Planetary Astronomy Fellow}
\altaffiltext{20}{Cape Breton University, 1250 Grand Lake Road, Sydney, NS B1P 6L2, Canada} 
\altaffiltext{21}{St. Mary's University, Halifax, NS B3H 3C3, Canada} 
\altaffiltext{22}{ Universit\"at Wien, Institut f\"ur Astronomie, T\"urkenschanzstrasse 17, A-1180 Wien, Austria } 
\altaffiltext{23}{Univ de Montr\'eal C.P. 6128, Succ. Centre-Ville, Montr\'eal, QC H3C 3J7, Canada } 
\altaffiltext{24}{Obs. du mont M\'egantic, Notre-Dame-des-Bois, QC J0B 2E0, Canada}
\altaffiltext{25}{SETI Institute, 189 Bernardo Avenue, Mountain View, CA 94043, USA}
\altaffiltext{26}{NASA Ames Research Center, Moffett Field, CA 94035, USA}
\altaffiltext{27}{University of Toronto, 50 St. George Street, Toronto, ON M5S 3H4, Canada}
\altaffiltext{28}{Centro de Astrof\'isica, Universidade do Porto, Rua das Estrelas, 4150-762 Porto, Portugal}
\altaffiltext{29}{Instituto de Astrof\'isica e Ci\^encias do Espa\c{c}o, Universidade do Porto, CAUP, Rua das Estrelas, PT4150-762 Porto, Portugal}
\altaffiltext{30}{INAF - Osservatorio Astrofisico di Torino, Via Osservatorio 20, I-10025 Pino Torinese, Italy}

\begin{abstract}
We report the first planet discovery from the two--wheeled \Kepler\ (K2) mission: \thisplanet. The host star \thisstar\ is a bright ($V = 10.1$, $K = 8.0$) K1--dwarf with high proper motion, and a parallax--based distance of $\distance \pm \udistance$~pc. Based on high-resolution optical spectroscopy, we find that the host star is metal-poor with \feh~$= \fe \pm \ufe$, and has a radius $R_\star = \rst \pm \urst$~\rsun\ and mass $M_\star = \mst \pm \umst$~\msun. The star was observed by the \Kepler\ spacecraft during its Two-Wheeled Concept Engineering Test in February 2014. During the 9 days of observations, K2 observed a single transit event. Using a new K2 photometric analysis technique we are able to correct small telescope drifts and recover the observed transit at high confidence, corresponding to a planetary radius of $R_p = \rpl \pm \urpl$~\rearth. Radial velocity observations with the HARPS-N spectrograph reveal a $\mpl \pm$ \umpl~\mearth\ planet in a 9.1 day orbit, consistent with the transit depth, duration, and ephemeris. Follow--up photometric measurements from the MOST satellite confirm the transit observed in the K2 photometry and provide a refined ephemeris, making \thisplanet\ amenable for future follow--up observations of this latest addition to the growing population of transiting super-Earths around nearby, bright stars.

\end{abstract}


\section{Introduction}

After four years of nearly continuous photometric monitoring and thousands of planet discoveries \citep[e.g.][]{borucki, howardrates, muirhead13, batalha13, barclay12, mortonswift}, the primary \Kepler\ Mission came to an end in in May 2013 with the failure of the second of four reaction wheels used to stabilize the spacecraft. Without at least three functioning reaction wheels, the spacecraft is unable to achieve the fine pointing necessary for high photometric precision on the original target field. However, an extended mission called K2 was enabled by pointing along the ecliptic plane and balancing the spacecraft against Solar radiation pressure to mitigate the instability caused by the failed reaction wheels. The recently extended K2 mission enables renewed opportunities for transit science on a new set of bright target stars, albeit with somewhat reduced photometric precision compared to the original \Kepler\ mission \citep{howell}. 

Searching for transiting exoplanets around bright, nearby stars is important because measuring the precise masses and radii of transiting planets allows for characterization of their interior structures and atmospheres \citep{charatmos, rogers, knutson, teske, kreidberg}. This is particularly desirable for planets with masses intermediate to those of the Earth and Uranus, commonly referred to as super--Earths, because no such planet exists in our Solar System \citep{valencia06}. However, while the radii of \Kepler\ planets are often measured to high precision \citep{ballard}, their masses are generally unknown because the host stars are faint ($V > 12$) and exposure times needed for radial velocity (RV) measurements are prohibitive for all but the brightest {\em Kepler} planet candidates \citep[e.g.][]{dumusque14, marcymasses}. 

Preparations for the extended two--wheeled \Kepler\ mission included a 9--day test of the new observing mode in February of 2014. After the data were released to the public, \citet[][hereafter VJ14]{vj14} presented a photometric reduction technique that accounts for the motion of the spacecraft, improves photometric precision of raw K2 data by a factor of 2--5, and enables photometric precision comparable to that of the original \Kepler\ Mission. 

While the data collected during the Engineering Test were intended primarily as a test of the new spacecraft operating mode, inspection of light curves produced with this technique nonetheless revealed a single transit event in engineering data taken of \thisstar. In this paper, we provide an analysis of that light curve along with archival and follow--up spectroscopy, archival and adaptive optics imaging, radial velocity measurements from the HARPS-N spectrograph, and photometric observations from the Wide Angle Search for Planets (WASP) survey and the Microvariability and Oscillations of STars (MOST) space telescope. These measurements allow us to verify and characterize the first planet discovered by the two-wheeled \Kepler\ mission, a new transiting super-Earth orbiting the bright, nearby, high-proper-motion K dwarf \thisstar. 

\section{Data and Analysis}

\subsection{K2 Photometry}

\begin{figure*}
\epsscale{1}
  \begin{center}
      \leavevmode
\plotone{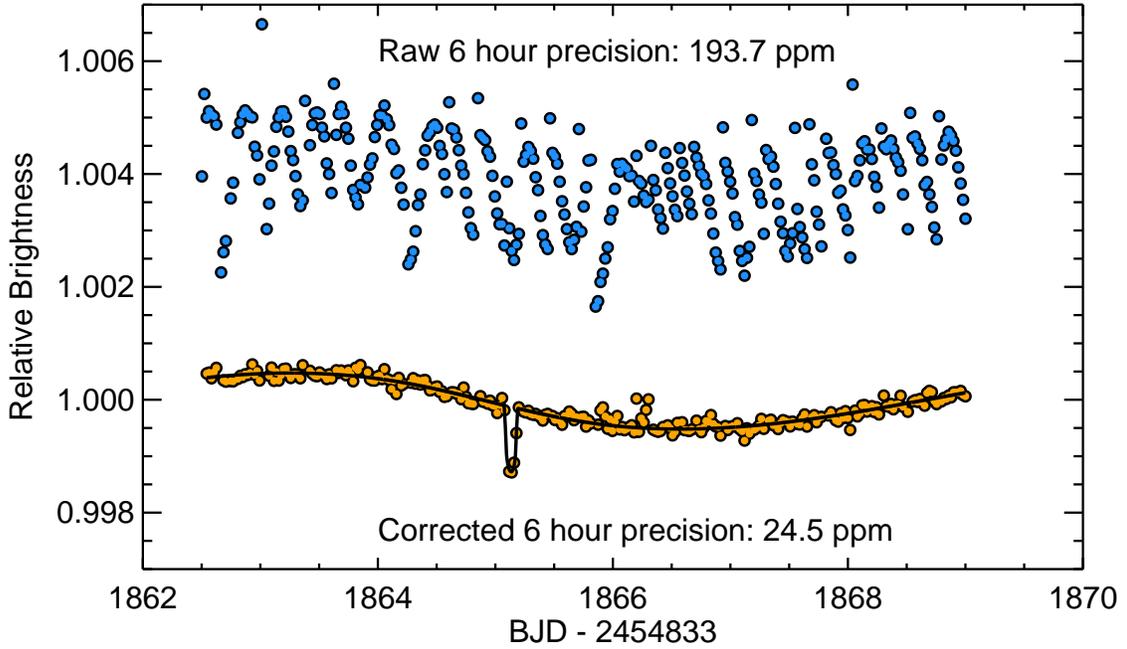}
\caption{Raw (top, blue) and SFF corrected (bottom, orange) K2 light curves. K2 only observed \thisstar\ for 9 days in February 2014, the last 6.5 of which are shown here. The raw data is vertically offset for clarity. A transit model light curve multiplied by a basis spline fit to the out-of-transit variations is overplotted on the corrected K2 data. The 6 hour photometric precision on this target (as defined by VJ14) improves by a factor of 7 as a result of the SFF processing.} \label{sff}
\end{center}
\end{figure*}

\thisstar\ and nearly 2000 other stars were observed by the \Kepler\ spacecraft from 4 February 2014 until 12 February 2014 during the \Kepler\ Two-Wheel Concept Engineering Test. After the first 2.5 days of the test, \Kepler\ underwent a large, intentional adjustment to its pointing to move its target stars to the center of their apertures, where they stayed for the last 6.5 days of the test. We downloaded the full engineering test dataset from the Mikulski Archive for Space Telescopes (MAST), and reduced the \Kepler\ target pixel files as described in VJ14. 

In brief, we extracted raw aperture photometry and image centroid positions from the \Kepler\ target pixel files. Raw K2 photometry is dominated by jagged events corresponding to the motion of the spacecraft, as \Kepler's pointing drifts due to Solar radiation pressure and is periodically corrected with thrusters. For the last 6.5 days after the large pointing tweak, we removed the systematics due to the motion of the spacecraft by correlating the measured flux with the image centroid positions measured from photometry. We essentially produced a ``self--flat--field'' (SFF) similar to those produced by, for instance, \citet{ballardspitzer}, for analysis of \Spitzer\ photometry. We fit a piecewise linear function to the measured dependence of flux on centroid position, with outlier exclusion to preserve transit events, and removed the dependence on centroid position from the raw light curve. Similarly to VJ14, we excluded datapoints taken while \Kepler's thrusters were firing from our reduced light curves because these data were typically outliers from the corrected light curves.  For \thisstar, the median absolute deviation (MAD) of the 30 minute long-cadence datapoints improved from $\simeq 500$ parts per million (ppm) for the raw light curve, to $\simeq$ 50 ppm for the SFF light curve. 

Visual inspection of light curves from the $\simeq$ 2000 targets observed during the Engineering Test revealed a one millimagnitude (mmag) deep candidate transit in photometry of \thisstar, designated EPIC\,60021410 by the \Kepler\ team. Raw and corrected K2 photometry for \thisstar\ are shown in Figure \ref{sff}. We fit a \citet{mandelagol} model to the transit and measured a total duration of approximately 2.25 hours, and a planet--to--star radius ratio of approximately 0.03.  Unfortunately, the data point during transit ingress happened during a thruster firing event, and was excluded by our pipeline. This particular data point does not appear to be anomalous, but we choose to exclude it to minimize risk of contaminating the transit with an outlier. Slow photometric variability, presumably due to starspot modulation, is evident in the K2 light curve at the sub-percent level. 

We also performed a similar SFF correction to the data taken in the 2.5 days of data before the large pointing tweak. Even though the resulting data quality is somewhat worse, we are able to confidently exclude any other events of a similar depth during that time. 

Because K2 only observed one transit event, we were not able to measure a precise orbital period for the planet candidate. Nonetheless, we were able to put rough constraints on the orbital period from our knowledge of the transit duration and estimates of the stellar properties.  The nine-day time baseline of the K2 observations allowed us to constrain the period of the candidate transiting planet to be greater than 5 days. To put a rough upper bound on the allowed planet period, we compared the transit duration of the candidate transit around \thisstar\ to the distribution of transit durations from the ensemble of \Kepler\ planet candidates \citep[retrieved from the NASA Exoplanet Archive;][]{nea}. We found that of the 413 \Kepler\ planet candidates with transit durations between 2 and 2.5 hours, 93\% had orbital periods shorter than 20 days. Because transit duration is a function of the mean stellar density, we repeated this calculation while restricting the sample to the 64 planet candidates with transit durations between 2 and 2.5 hours and host star effective temperatures within 200~K of \thisstar. We find that similarly, 94\% of these candidates had orbital periods shorter than 20 days. 

\begin{deluxetable}{lrcl}
\tablecaption{Astrometric and Photometric Properties of \thisstar
\label{tab:kicphot}}
\tablewidth{0pt}
\tablehead{
\colhead{Parameter} & \colhead{Value} & \colhead{Uncertainty} & \colhead{Source}  
}
\startdata
$\alpha$ (J2000)  & 23\,35\,49.28 & ... & \Hipparcos\tablenotemark{a} \\
$\delta$ (J2000)  & +00\,26\,43.86 & ... & \Hipparcos \\
$\mu_\alpha$ (mas yr$^{-1}$) & -238.0 & 1.7 & \Hipparcos \\
$\mu_\delta$ (mas yr$^{-1}$) & -185.9 & 0.9 & \Hipparcos \\
$\pi$ (mas) & 18.1 & 1.72 & \Hipparcos \\

$B$ & 11.08 & 0.01 & \Tycho \\
$V$ & 10.190 &  0.009 & \Tycho\tablenotemark{b} \\
$R$ & 9.71 & 0.03 & TASS\tablenotemark{c} \\
$I$ & 9.25 & 0.03 & TASS \\
$u$ & 14.786 & 0.02 & SDSS\tablenotemark{d} \\
$g$ & 10.837 & 0.02 & SDSS \\
$r$ & 9.908  & 0.02 & SDSS \\
$i$ & 9.680  & 0.02 & SDSS \\
$J$        & 8.60 & 0.02 & 2MASS\tablenotemark{e} \\
$H$        & 8.14 & 0.03 & 2MASS \\
$K_S$      & 8.03 & 0.02 & 2MASS
\enddata
\tablenotetext{a}{\citet{hipparcos}}
\tablenotetext{b}{\citet{tycho}}
\tablenotetext{c}{\citet{tass}}
\tablenotetext{d}{\citet{sdss}}
\tablenotetext{e}{\citet{2mass}}
\end{deluxetable}

\subsection{Imaging}
\subsubsection{Archival Imaging}

\begin{figure*}[t!]
\epsscale{1}
  \begin{center}
      \leavevmode
\plotone{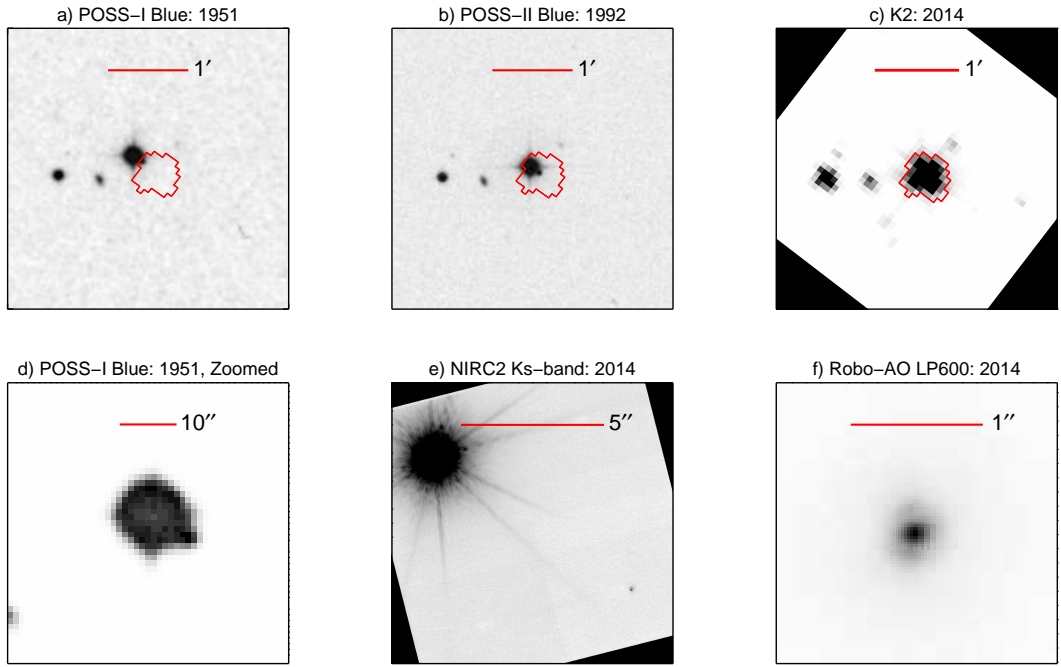}
\caption{Imaging of \thisstar. Archival image from the POSS-I survey taken in 1951, showing a clear background in the K2 aperture (shown in red). b) Archival image from the POSS-II survey, taken in 1992, showing the high proper motion of the star. c) Co-added image of the last 5 days of the K2 Engineering test. d) Zoomed and scaled version of the POSS-I image showing the companion. e) Modern Keck/NIRC2 image of the \thisstar\ system showing that the companion shares proper motion with \thisstar. In this image, the primary was intentionally saturated to simultaneously image the companion. f) Robo-AO adaptive optics image in an optical bandpass close to that of \Kepler, showing no apparent close companions. NIRC2 images also exclude companions at even closer angular separations, but in infrared bandpasses. } \label{imaging}
\end{center}
\end{figure*}

We used a combination of modern and archival imaging to limit potential false--positive scenarios for the transit event. \thisstar\ was observed in the National Geographic Society--Palomar Observatory Sky Survey (POSS-I) on 28 November 1951. \thisstar\ has a proper motion of 303~mas~yr$^{-1}$ \citep{hipparcos}, and therefore has moved nearly 20 arcseconds with respect to background sources since being imaged in POSS-I. Inspection of the POSS-I image reveals no background objects within the K2 aperture used in our photometric reduction. We show the POSS-I blue image overlaid with the K2 photometric aperture in Figure \ref{imaging}a. The POSS-I survey has a limiting magnitude of 21.1 in the blue bandpass \citep{abell}, 10 magnitudes fainter than \thisstar. The depth of the detected transit is 0.1\%, so if a background eclipsing binary were responsible the depth would correspond to a total eclipse of a star 7.5 magnitudes fainter. A low--proper--motion background star such as that would have readily been detected in POSS-I imaging. We conclude that our aperture is free of background objects whose eclipses could masquerade as planet transits. 

The POSS-I imaging also reveals a companion star about 8 arcseconds to the southwest of \thisstar. The companion is not fully resolved in POSS-I because the photographic plate was saturated by the bright primary star, but an asymmetry in the stellar image is visible. \thisstar\ was also observed during the Second Palomar Observatory Sky Survey (POSS-II) on 31 August 1992. Improvements in photographic plate technology over the previous forty years allowed the companion star to be resolved. The companion shares a common proper motion with the primary at a projected distance of $\simeq 500$~AU, so we conclude the two stars are a gravitationally--bound visual binary system.

\subsubsection{Modern imaging}

\thisstar\ was observed during the Sloan Digital Sky Survey (SDSS), and the secondary star was detected \citep{sdss}. The secondary star falls on a diffraction spike caused by the much brighter primary star, but the SDSS pipeline flagged its photometry as ``Acceptable.'' The SDSS photometry indicates that the secondary star is 6--7 magnitudes fainter than the primary depending on the filter, so because the two stars are gravitationally associated, the secondary must be intrinsically much fainter than the K--dwarf primary. This implies that the companion must either be a late M--dwarf or a white dwarf. The SDSS colors are relatively flat, indicating a hot star. To quantify this, we fit the {\em ugri} SDSS colors to a blackbody model, excluding {\em z} due to its low throughput and assuming photometric errors of 5\%. We included no corrections for extinction due to the proximity of the target and our ability to accurately predict broadband photometry using stellar models in Section \ref{stellarproperties}. We find that the data are best described by an object radiating at a temperature of $T_{\rm WD} = 7500 \pm 200$~K. We used the Stefan--Boltzmann law combined with the \Hipparcos\ parallax and derived temperature to estimate a radius of $R_{\rm WD} = 1.2 \pm 0.1 R_\earth$, which is consistent with our white dwarf hypothesis. Using a simple analytic white dwarf cooling law \citep{mestel, wdcooling}, we estimate a cooling age of the white dwarf of $t_{\rm cool} \sim 1.3$ Gyr. The formal uncertainty on the cooling age is 0.2 Gyr, but this neglects uncertainties due to the unknown composition of the white dwarf and inaccuracies in the simple model. The true uncertainty on this quantity is likely on order of a factor of two \citep{vanhorn}. The cooling age of the white dwarf is a lower limit on the age of the system, and the total age of the system is the sum of the main sequence lifetime of the progenitor and the white dwarf's cooling age.   

The secondary star is close enough to the primary that it is blended in the K2 image and is bright enough that if it were a totally eclipsing binary, it could cause the transit event we observed. This situation is unlikely because the duration and minimum period of the event are generally inconsistent with an object eclipsing a white dwarf. With the baseline of K2 data we can exclude orbital periods shorter than 5 days. While 5--day period companions eclipsing main sequence stars are common, and have relatively high transit probabilities, the probability of a transit or eclipse goes as $P \propto R_\star$ at a given stellar mass and orbital period. Furthermore, in order for an Earth sized object eclipsing a white dwarf to have an eclipse duration of two hours, the orbital period would have to be roughly 600 years in the case of a circular orbit and impact parameter $b = 0$. Even with a highly elliptical orbit transiting at apastron, which is {\em a priori} unlikely, the orbital period would be of order centuries, and the semimajor axis would be roughly 50~AU. The probability of an orbit such as that eclipsing the white dwarf is $P \sim (R_{\star} + R_{p})/a \sim 10^{-6}$, where $a$ is the semimajor axis and $R_{p}$ is the radius of the occulting body. In the worst--case scenario of a non-luminous Jupiter sized object occulting the white dwarf, the orbital period would have to be on order 3 years and have a semimajor axis of roughly 1.5 AU, correpsonding to a transit probability of $P \sim 10^{-4}$. We conclude that the transit event we observed was far more likely caused by a short--period planet orbiting the primary star than a long period object eclipsing the secondary.    

\subsubsection{Adaptive optics imaging}

We also obtained high--angular--resolution imaging of the primary star to rule out any very close associated companions. We observed \thisstar\ with the Robo-AO laser adaptive optics and imaging system on the 60--inch telescope at Palomar Observatory \citep{baranec, lawroboao}. We obtained seven images with Robo-AO between 15 June 2014 and 11 July 2014 in three different bandpasses: Sloan $i$--band, Sloan $z$--band, and a long--pass filter with a cutoff at 600~nm (LP600) which more closely approximates the \Kepler\ bandpass. Each observation consisted of a series of images read out from the detector at a maximum rate of 8.6~Hz, for a total integration time of 90 seconds. The frames were co--added in post-processing using a shift and add technique with \thisstar\ as the tip--tilt guide star. 

The quality of the Robo-AO images varied between the observations, but none of the images showed evidence for companions within 3 magnitudes of the primary outside of 0.2 arcseconds. Some but not all of the images, however, showed an elongation that could be consistent with a bright close binary companion at a separation of 0.15 arcseconds at the \textless 5--$\sigma$ level, similar to KOI 1962 in \citet{lawroboao}. 

To investigate this possibility further, we obtained higher resolution adaptive optics images on 2 August 2014 using the Keck II Natural Guide Star Adaptive Optics (NGSAO) system with the NIRC2 narrow detector at Keck Observatory. We obtained unsaturated frames of \thisstar\ in $J$, $H$, and $K_S$-band filters to search for close companions near the diffraction limit ($\sim$40~mas in $H$-band).  We also acquired deeper, saturated images in $H$ (70~sec total) and $K_S$-bands (200 sec total) with the primary positioned in the lower right quadrant of the array and rotated so the white dwarf companion falls in the field of view. We calibrated and processed the data as described in \citet{bowler}. We corrected the data for optical aberrations using the distortion solution from B. Cameron (2007, priv. communication) and North-aligned the images using the detector orientation measured by \citet{yelda}. We found no evidence for the companion suggested by some of the Robo-AO data. Our 7-$\sigma$ $H$-band limiting contrasts are \{3.0, 5.9, 6.8, 9.2, 10.8, 12.7\}~mag at separations of \{0$\farcs$1, 0$\farcs$3, 0$\farcs$5, 1$\farcs$0, 2$\farcs$0, 5$\farcs$0\}. We are able to exclude roughly equal brightness companions to an angular separation of 0.04 arcseconds (projected distance of 2.2 AU). 

\subsection{Reconnaissance spectroscopy} \label{spectroscopy}

\thisstar\ was observed 9 times for the Carney-Latham Proper Motion Survey \citep{carneylatham} with the CfA Digital Speedometer spectrograph over the course of 9.1 years from 1982 until 1991. The Digital Speedometer measured radial velocities to a precision of approximately 0.3~\kms, and detected no significant radial velocity variations or trends in the velocities of \thisstar. When corrected into an absolute radial velocity frame, the Digital Speedometer measurements indicate an absolute radial velocity of $-3.06 \pm 0.12$~\kms\ and when combined with proper motion, a space velocity of $(U,V,W) = (-86.7,-0.2,4.5) \pm (7.6, 1.2,0.5)$~\kms. This somewhat unusual space velocity corresponds to an elliptical orbit in the plane of the galaxy, indicating that \thisstar\ originated far from the stellar neighborhood. A detailed analysis of \thisstar's elemental abundances could reveal patterns that are dissimilar to stars in the Solar neighborhood.

We obtained three observations of \thisstar\ in June of 2014 with the Tillinghast Reflector Echelle Spectrograph (TRES) on the 1.5 m Tillinghast Reflector at the Fred L. Whipple Observatory.  The spectra were taken with a resolving power of $R = 44,000$ with a signal--to--noise ratio (SNR) of approximately 50 per resolution element. When corrected into an absolute radial velocity frame, the TRES spectra indicate an absolute radial velocity for \thisstar\ of $-3.12 \pm 0.1$~\kms. When combined with the absolute radial velocities from the Digital Speedometer, there is no evidence for radial velocity variation of greater than 100~\ms\ over the course of 30 years. The three individual radial velocities from the TRES spectra revealed no variability at the level of 20~\ms\ over the course of eight days.  We also find no evidence for a second set of stellar lines in the cross correlation function used to measure the radial velocities, which rules out many possible close companions or background stars. When the adaptive optics constraints are combined with a lack of radial velocity variability of more than 100 m/s over 30 years and the lack of a second set of spectral lines in the cross correlation function, we can effectively exclude any close stellar companions to \thisstar.

\subsection{HARPS-N Spectroscopy}

\begin{figure}[t!]
\epsscale{1}
  \begin{center}
      \leavevmode
\plotone{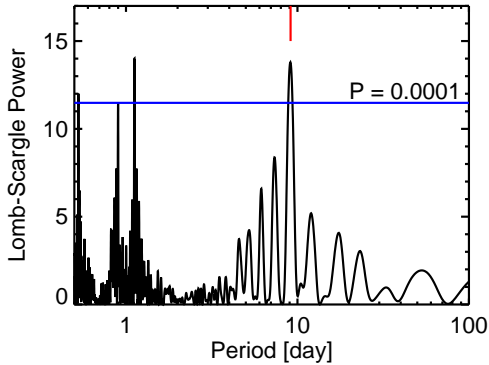}
\caption{Left: Lomb-Scargle periodogram of the HARPS-N radial velocity data. We find a strong peak at a period of 9.1 days and see daily aliases of the 9.1 day signal with periods close to one day. The horizontal blue line indicates a false alarm probability of 0.0001, and the vertical red hash mark indicate the period (9.12 days) from our combined analysis described in Section \ref{joint}. } \label{periodogram}
\end{center}
\end{figure}

We obtained 44 spectra of \thisstar\ on 33 different nights between July and October of 2014 with the HARPS-N spectrograph \citep{harpsn} on the 3.57m Telescopio Nazionale Galileo (TNG) on La Palma Island, Spain to measure precise radial velocities and determine the orbit and mass of the transiting planet. Each HARPS-N spectrum was taken with a resolving power of $R = 115000$, and each measurement consisted of a 15 minute exposure, yielding a SNR of 50--100 per resolution element at 550~nm, depending on weather conditions. The corresponding (formal) radial velocity precision ranged from 0.90~\ms\ to 2.35~\ms. Radial velocities were extracted by calculating the weighted cross correlation function of the spectrum with a binary mask \citep{xcor, wxcor}. In some cases, we took one 15 minute exposure per night, and in other cases, we took two 15 minute exposures back-to-back. In the latter case we measured the two consecutive radial velocities individually and report the average value. 

\begin{figure*}[t!]
\epsscale{1}
  \begin{center}
      \leavevmode
\plotone{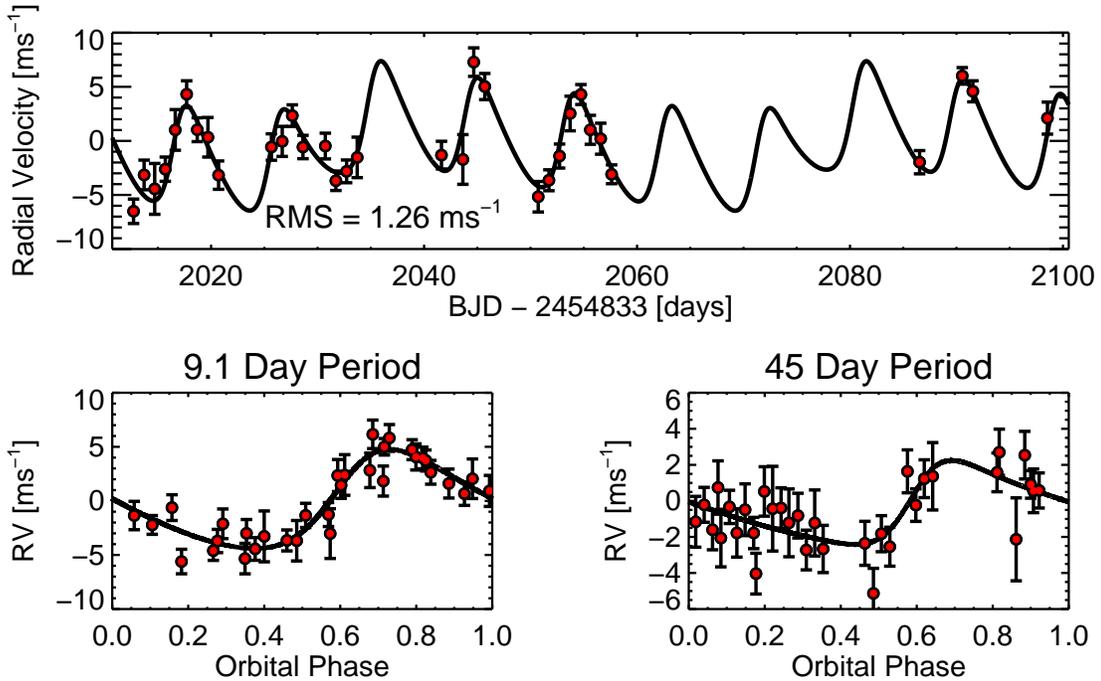}
\caption{Top: All radial velocity measurements of \thisstar, with observations taken during the same night binned together. We strongly detect a 9.1 day periodicity and find more tenuous evidence for a 45 day periodicity. Bottom left: RV measurements phase folded on the 9.1 day period with the best--fit 45 day signal removed. Bottom right: RV measurements phase folded on the 45 day period with the best--fit 9.1 day signal removed.} \label{rvplot}
\end{center}
\end{figure*}

The HARPS-N radial velocity measurements are listed in Table \ref{rvs}, and are shown in Figure \ref{rvplot}. A periodic radial velocity variation with a period of about nine days and a semiamplitude of about 4~\ms\ is evident in the RV time series. We checked that we identified the correct periodicity by calculating a Lomb-Scargle periodogram \citep{scargle}, shown in Figure \ref{periodogram}. We found a strong peak at a period of 9.1 days and an aliased peak of similar strength with a period close to 1 day, corresponding to the daily sampling alias of the 9.1 day signal \citep[e.g.][]{dawsonfabrycky}. We estimated the false alarm probability of the RV detection by scrambling the RV data and recalculating the periodogram numerous times, and counting which fraction of the scrambled periodograms have periods with higher power than the unscrambled periodogram. We found that the false alarm probability of the 9.1 day periodicity is significantly less than $10^{-4}$.

In addition to the 9.1 day signal, we also found evidence for a weaker 45 day periodic RV variation.To help decide whether to include the second periodicity in our RV modeling, we fit the HARPS-N radial velocities with both a one-planet and a two-planet Keplerian model. The one planet model was a Keplerian function parameterized by: $\log({P})$, time of transit, $\log{(\rm RV~ semiamplitude)}$,$\sqrt{e}\sin(\omega)$, $\sqrt{e}\cos(\omega)$, where $P$ is the planet's orbital period, $e$ is the orbital eccentricity, and $\omega$ is the argument of periastron. We also fit for a radial velocity zero-point and a stellar jitter term, for a total of 7 free parameters. We fit each of these parameters with an unbounded uniform prior except for $\sqrt{e}\sin(\omega)$ and $\sqrt{e}\cos(\omega)$, which had uniform priors over the interval between -1 and 1. The two planet model was the sum of two Keplerian functions, each of which was parameterized by: $\log({P})$, time of transit, $\log{(\rm RV~semiamplitude)}$,$\sqrt{e}\sin(\omega)$, and $\sqrt{e}\cos(\omega)$. Once again, we also fit for a radial velocity zero point and stellar jitter term, for a total of 12 free parameters. We fit each of these parameters with an unbounded uniform prior except for $\sqrt{e}\sin(\omega)$ and $\sqrt{e}\cos(\omega)$, which had uniform priors over the interval between -1 and 1, and for $\log{P_2}$, the period of the outer planet, which we constrained to be between the period of the inner planet and 1000 days. We performed the fits using \emcee\ \citep{emcee}, a Markov Chain Monte Carlo (MCMC) algorithm with an affine invariant ensemble sampler. We note that upon exploring various different periods for the outer planet, our MCMC analysis found the 45 day period to be optimal. We calculated the Bayesian Information Criterion \citep[BIC,][]{bic} to estimate the relative likelihoods of the two models. While the BIC does not provide a definitive or exact comparison of the fully marginalized likelihoods of the models, it allows us to roughly estimate the relative likelihoods. Upon calculating the BIC, we estimate that the two-planet model is favored over the one-planet model with confidence $P\sim0.03$. From here on, we therefore model the radial velocity observations as the sum of two Keplerian functions.

\begin{figure*}[t!]
\epsscale{1}
  \begin{center}
      \leavevmode
\plotone{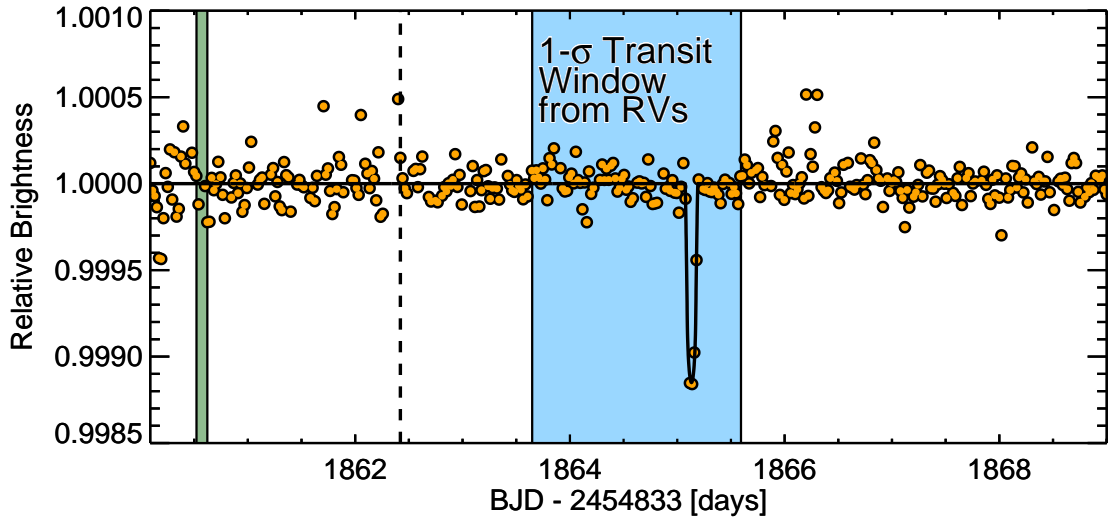}
\caption{K2 light curve (orange dots) overlaid with the transit window derived from the HARPS-N radial velocities (blue shaded region). We show the full K2 light curve, including data taken before \Kepler's large pointing tweak at t$\simeq$ 1862.4 days (data to the left of the dashed line). The green shaded region near t$\simeq$ 1861.6 days is half a phase away from the transit, assuming a 9.12 day period. No secondary eclipse is visible, lending credence to the planetary interpretation of the transit and RV variations.} \label{transitwindow}
\end{center}
\end{figure*}

For both periods, we find an amplitude consistent with that of a transiting super-Earth. The nine-day periodicity in the RVs is consistent with the orbital period we estimated from the duration of the K2 transit event. We ``predicted'' the time of transit for the nine day period planet during the K2 observations, and found that the HARPS-N measurements alone constrain the expected time of transit to better than one day, and we find that the K2 transit event is consistent with the transit ephemeris predicted by only the HARPS-N RVs at the 68.3\% (1-$\sigma$) level. We show the K2 light curve with the transit window derived from only the HARPS-N RVs in Figure \ref{transitwindow}. Thus, the 9.1 day periodicity is consistent with being caused by a transiting planet. The more tenuous 45 day periodic variation, on the other hand, may be due to to an outer planet, but it may also be caused by stellar variability.  

The HARPS-N spectra include regions used to calculate activity indicators such as the Mount Wilson S$_{\rm HK}$ index and the R$^\prime_{\rm HK}$ index \citep[e.g.][]{wright04}. We calculated the S$_{\rm HK}$ index for each HARPS-N spectrum and found a mean value of 0.275 $\pm$ 0.0034, and an associated $\log_{10}{R'_{\rm HK}}= -4.773 ~\pm~ 0.007$. There were no obvious correlations between the S$_{\rm HK}$ index and either the measured RV or the residuals to either a one or two planet Keplerian fit. 

\begin{deluxetable}{lrc}
\tablewidth{0pt}
\tablecaption{ HARPS-N Radial Velocities of \thisstar\ \label{rvs}}
\tablehead{
\colhead{BJD - 2454833} & \colhead{RV [\ms]} &\colhead{$\sigma_{\rm RV}$ [\ms]} }
\startdata
2012.7150 & -6.51 & 1.13 \\
2013.7062 & -3.15 & 1.37 \\
2014.7001 & -4.44 & 2.35 \\
2015.6955 & -2.62 & 1.12 \\
2016.6307 & 1.01 & 1.88 \\
2017.7029 & 4.31 & 1.24 \\
2018.6971 & 1.03 & 1.10 \\
2019.6985 & 0.33 & 1.83 \\
2020.7000 & -3.17 & 1.31 \\
2025.6645 & -0.57 & 1.22 \\
2026.6780 & -0.05 & 1.39 \\
2027.6258 & 2.33 & 1.00 \\
2028.6266 & -0.57 & 1.08 \\
2030.7261 & -0.48 & 1.19 \\
2031.7186 & -3.69 & 0.90 \\
2032.7231 & -2.82 & 1.05 \\
2033.7197 & -1.53 & 1.87 \\
2041.6353 & -1.32 & 1.28 \\
2043.6442 & -1.72 & 2.30 \\
2044.6658 & 7.28 & 1.32 \\
2045.7016 & 5.02 & 1.21 \\
2050.7205 & -5.16 & 1.41 \\
2051.7258 & -3.64 & 0.97 \\
2052.7229 & -1.40 & 1.11 \\
2053.7166 & 2.53 & 1.60 \\
2054.7304 & 4.29 & 0.92 \\
2055.6129 & 1.02 & 1.34 \\
2056.5885 & 0.20 & 1.43 \\
2057.6126 & -3.09 & 0.87 \\
2086.5397 & -1.97 & 1.07 \\
2090.5445 & 6.01 & 0.76 \\
2091.5401 & 4.58 & 0.97 \\
2098.5448 & 2.10 & 1.48 \\

\enddata
\end{deluxetable}

\subsection{Photometry}

\subsubsection{WASP}

\thisstar\ was observed by the SuperWASP-N instrument on La Palma, Spain and the SuperWASP-S instrument at the South African Astronomical Observatory \citep{wasp}. The observations spanned three observing seasons, from 2008 to 2010, and consisted of roughly 15,000 data points with a typical precision of 0.6\%. The WASP observations are consistent with a typical level of stellar variability at the sub-percent level. The WASP data rule out deep transits but are not of high enough quality to detect the 0.1\% transit depth observed by K2. In Section \ref{prot}, we use the WASP light curve in combination with light curves from K2 and MOST to attempt to derive \thisstar's rotation period.

\subsubsection{MOST}

After detecting the K2 transit, we obtained follow--up photometric observations with the Microvariability and Oscillations of STars \citep[MOST,][]{most} space telescope, during August and September of 2014. MOST observed \thisstar\ during three nearly continuous time spans: 13 days from 3 August 2014 to 16 August 2014, 18 days from 21 August 2014 to 9 September 2014, and 3.5 days from 15 September 2014 to 18 September 2014. During the first segment of the MOST data, observations of \thisstar\ were interleaved with observations of other stars during the satellite's orbit around Earth, but for the second and third segments MOST observed \thisstar\ continuously. During the first and third segments, the exposure time for individual datapoints was 1 second, and during the second segment, the exposure time was 2 seconds.

We processed the MOST data using aperture photometry techniques as described in \citet{mostdata}. Background scattered light (modulated by the 101 minute orbital period of MOST) and interpixel sensitivity variations were removed by subtracting polynomial fits to the correlations between the stellar flux, the estimated background flux, and the centroid coordinates of the star. At each stage, outlying data points were excluded by either sigma--clipping or hard cuts. The resulting precision of the MOST light curve was approximately 0.2\% per 2--second exposure. 

When we search the MOST light curve at the predicted times of transits from a simultaneous analysis of the K2 and HARPS-N data, we detect a weak signal with the same depth, duration, and ephemeris as the K2 transit. The MOST data refine our estimate of the transiting planet period to a precision of roughly 30 seconds. We take this detection as confirmation that the 9.1 day period detected in radial velocities is in fact caused by the transiting planet. From here on, we refer to the 9.1 day period planet as \thisplanet. 

\begin{figure}
\epsscale{1}
  \begin{center}
      \leavevmode
\plotone{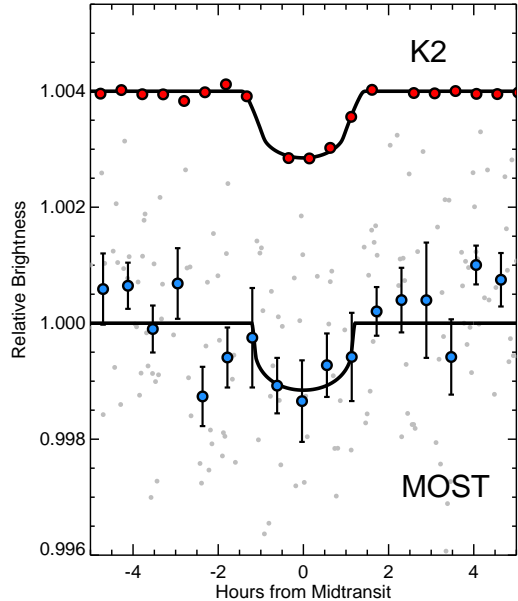}
\caption{K2 light curve (red dots) and binned MOST light curves (blue dots). Best fit models are overplotted in solid black lines. Individual MOST datapoints are shown as grey dots. The K2 light curve is vertically offset for clarity. The MOST data yield a marginal ($\simeq$ 3--$\sigma$)detection of the transit at the time predicted by HARPS-N radial velocities and the K2 light curve.} \label{k2most}
\end{center}
\end{figure}
\section{Analysis and Results}

\subsection{Stellar Properties}
\subsubsection{Spectroscopic Parameters of the Primary}

We measured the spectroscopic properties of the host star using the Stellar Parameter Classification (SPC) method \citep{spc} on the spectra from TRES and HARPS-N. Analysis of spectra from both instruments showed consistent results for the stellar parameters. We adopt the results from the HARPS-N spectra, due to their higher spectral resolution and SNR. The SPC analysis yields an effective temperature of $5089 \pm 50$~K, a metallicity [M/H]~$ = -0.16 \pm 0.08$, and a surface gravity $\log{g_\star} = 4.55 \pm 0.1$. We did not detect significant rotational broadening even with the high resolution HARPS-N spectra. The upper limit on the projected rotational velocity is roughly $v\sin(i) \lesssim 2$~km~s$^{-1}$. 

\subsubsection{Stellar Mass and Radius} \label{stellarproperties}

We used several different approaches to estimate the stellar mass and radius of \thisstar. First, we used the SPC parameters, in particular the metallicity, surface gravity, and effective temperature, to interpolate onto the Yonsei--Yale stellar evolution model grids \citep{yy} using a Monte Carlo approach. The resulting stellar parameters were $M_\star = 0.772 \pm 0.033 M_\sun$, and $R_\star = 0.746 \pm 0.042 R_\sun$. 

\thisstar\ was observed by \Hipparcos\ and has a measured parallax, allowing us to interpolate model grids using a separate luminosity indicator. We used the online Padova model interpolator\footnote{\url{http://stev.oapd.inaf.it/cgi-bin/param}}, which uses the technique of \citet{dasilva} to interpolate a measured effective temperature, metallicity, $V$--band magnitude, and parallax onto PARSEC isochrones, yielding estimates of stellar mass, radius, surface gravity, and $\bv$ color. When the SPC parameters, \Hipparcos\ parallax, and the V-- magnitude from the \Tycho\ catalog \citep{tycho} are provided as input, the models output $M_\star = \mst \pm \umst M_\sun$, and $R_\star = \rst \pm \urst R_\sun$, along with $\log{g_\star} = \loggst \pm \uloggst$ dex and $\bv = 0.935 \pm 0.018$ mag. 

The model--predicted surface gravity is consistent with the spectroscopically measured surface gravity and more precise due to the additional constraint from parallax. The model output $\bv$ color is discrepant with the measured \Tycho\ $\bv$ at the 1.5--$\sigma$ level, but is still within 0.04 magnitudes of the measured \Tycho\ $\bv = 0.9$. This discrepancy is small enough (3\%) that it could be due to differences in the filters used by \Tycho\ and the filter transmission assumed by the Padova models. 

We adopt the outputs from the Padova model interpolator as our stellar parameters due to their more precise constraints and ability to predict $\log{g_\star}$ and $\bv$. 

\subsubsection{Stellar Rotation Period}\label{prot}
\begin{figure*}
\epsscale{1}
  \begin{center}
      \leavevmode
\plotone{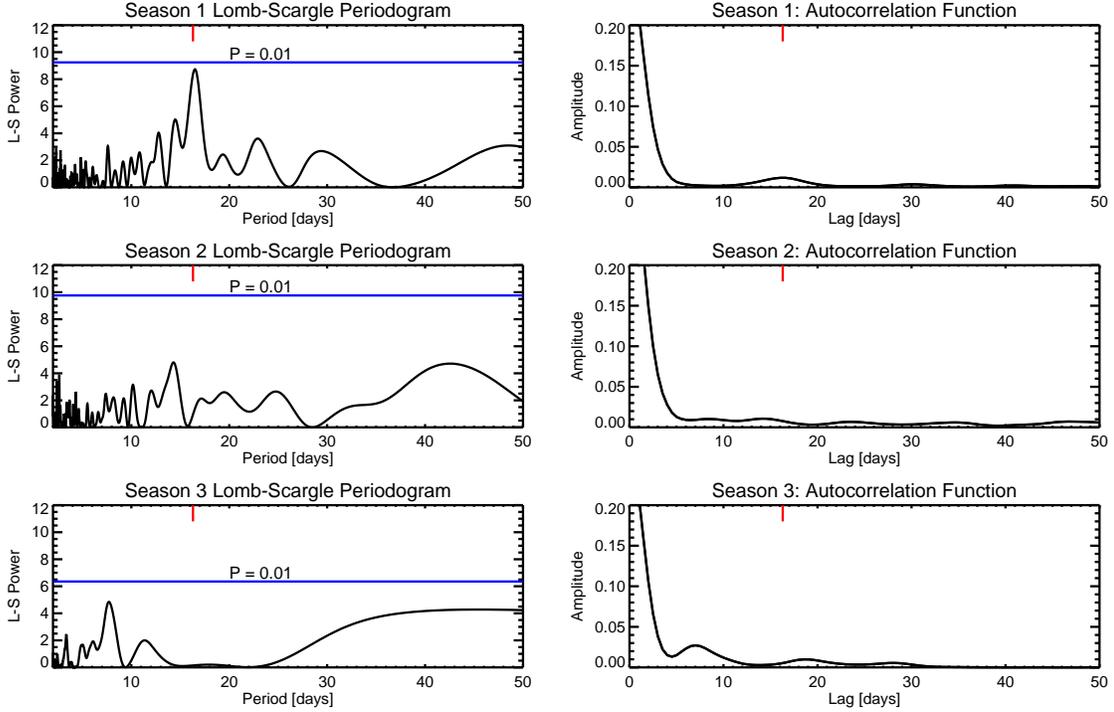}
\caption{Lomb-Scargle periodograms and autocorrelation functions for three observing seasons of WASP data. The first observing season (2008) shows evidence for a roughly 16 day rotation period in both the periodogram and autocorrelation function. We mark a 16.3 day period in each figure with a red hash mark, and show the level of a 1\% false alarm probability with a horizontal blue line. } \label{rotationperiod}
\end{center}
\end{figure*}

We attempted to measure the rotation period of \thisstar\ using photometric measurements from WASP, MOST, and K2 to see if stellar activity might contribute to the possible 45 day periodicity in the RV measurements. A constraint or measurement of the rotation period consistent with the possible 45 day periodicity in radial velocities could affect our interpretation of the signal. We only used photometric measurements for our analysis given the relatively short time coverage and sparseness of the spectroscopic observations.

We first started by analyzing the WASP data only, because its time baseline far exceeded that of the K2 and MOST data. We binned the WASP data into nightly datapoints, calculated a Lomb-Scargle periodogram and Fourier transformed the resulting power spectrum to obtain an autocorrelation function. We performed this analysis on each season of WASP data individually. In the first season (2008) of WASP data, we found a moderately strong peak in both the autocorrelation function and the Lomb-Scargle periodogram at a period of about 16 days. We evaluated the significance of this peak by scrambling the binned data, recalculating the Lomb-Scargle periodograms, and counting the number of times the maximum resulting power was greater than the power in the 16 day peak. We found a false alarm probability of 2\% for the peak in the first season. We did not find any convincing signals in the second (2009) or third (2010) observing seasons. A possible explanation for the inconsistency between observing seasons is that \thisstar\ experienced different levels of starspot activity, but it is also possible that the 16 day period detected in the first season is spurious. We concluded that the WASP data showed a candidate rotation period at 16 days, but the relatively high false alarm probability and lack of consistency between observing seasons precluded a confident detection.

After our analysis of the WASP data yielded suggestive yet ambiguous results, we attempted to measure the rotation period of \thisstar\ by fitting all of the photometric data with a Gaussian process noise model. Stochastic yet time--correlated processes such as the variability produced by rotation in stellar light curves can be modeled as a Gaussian process by parametrizing the covariance matrix describing the time--correlation between data points with a kernel function and inferring kernel hyperparameters from the data. We use a quasi--periodic kernel function for the specific problem of measuring a rotation period from a light curve. This in principle is better suited to inferring the rotation period of a star than a periodogram analysis because the variability produced by active surface regions on a rotating star is typically neither sinusoidal nor strictly periodic. The Gaussian process analysis also allows us to simultaneously model multiple datasets, and to take advantage of datasets (like K2 and MOST) with relatively short time coverage. 

We conducted our analysis using {\it george} \citep{Foreman-Mackey}, a Gaussian process library that employs Hierarchical Off-Diagonal Low Rank Systems (HODLR), a fast matrix inversion method \citep{Ambikasaran2014}. We used the following kernel function in our analysis:

\begin{equation}\label{kernel}
k_{ij} = A^2\exp\left[{\frac{-(x_i-x_j)^2}{2l^2}}\right]\exp\left[\frac{-\sin^2\left(\frac{\pi(x_i-x_j)}{P}\right)}{g_q^2}\right],
\end{equation}

\noindent where $A$ is the amplitude of correlation, $l$ is the timescale of exponential decay, $g_q$ is a scaling factor for the exponentiated sinusoidal term and $P$ is the rotation period. An additional hyperparameter, $s$ was used to account for additional white noise in the time series, where $s$ is added to the individual measurement uncertainties in quadrature. 

We modeled the three continuous periods of MOST data, the three seasons of WASP data, and the K2 photometry simultaneously, with $A$, $l$, $g_q$ and $P$ constrained to be the same across all seven data sets and with $s$ allowed to take a different value for each. We used \emcee\ to explore the posterior distributions of the hyperparameters. The resulting posterior distribution was not well constrained, with significant power at essentially all rotation periods greater than about 8 days and less than about 50 days. There were a few periods which seemed to be preferred to some extent in the posterior distribution -- a strong peak at 12 days, and weaker peaks at 16, 20, and 32 days. 

We conclude that with our data, we cannot conclusively identify a rotation period for \thisstar, which leaves us unable to rule out stellar activity as the cause of the 45 day signal in the radial velocities. While we do not find any strong evidence in the photometry that the rotation period is close to 45 days, we cannot conclusively rule out a 45 day rotation period. More photometric (or spectroscopic) observations will be important to determining \thisstar's rotation period.

\subsection{Joint Analysis and Planet Properties}\label{joint}
\begin{deluxetable*}{lcccc}
\tablecaption{System Parameters for \thisstar \label{tab:bigtable}}
\tablewidth{0pt}
\tablehead{
  \colhead{Parameter} & 
  \colhead{Value}     &
  \colhead{} &
  \colhead{68.3\% Confidence}     &
  \colhead{Comment}   \\
  \colhead{} & 
  \colhead{}     &
  \colhead{} &
  \colhead{Interval Width}     &
  \colhead{}  
}
\startdata
\emph{Orbital Parameters} & & & \\
Orbital Period, $P$~[days] & \perpl & $\pm$&$ \uperpl $ & A \\
Radius Ratio, $(R_P/R_\star)$ & \rprst & $\pm$&$ \urprst$ & A \\
Transit Depth, $(R_P/R_\star)^2$ & \depth & $\pm$&$ \udepth$ & A \\
Scaled semimajor axis, $a/R_\star$  & \arst & $\pm$&$ \uarst$ & A \\
Orbital inclination, $i$~[deg] & \incl & $\pm$&$ \uincl$ & A \\
Transit impact parameter, $b$ & \imp & $\pm$&$ \uimp$ & A \\
Eccentricity  & \ecc & $\pm$ & $\uecc$ & A \\
Argument of Periastron $\omega$~[degrees] & \omegap & $\pm$&$ \uomegap$ & A \\
Velocity semiamplitude $K_\star$~[m s$^{-1}$] &  \semi &$ \pm$&$ \usemi $ & A \\
Time of Transit $t_{t}$~[BJD] & \ttransit & $\pm$& \uttransit & A\\ 
 & & \\
\emph{Stellar Parameters} & & & \\
$M_\star$~[$M_\odot$] & \mst & $\pm$&$ \umst$ & B,D \\
$R_\star$~[$R_\odot$] & \rst & $\pm$&$ \urst$ & B,D \\
$\rho_\star$~[$\rho_\odot$] & \rhost  & $\pm$&$ \urhost$ & B,D \\
$\log g_\star$~[cgs] & \loggst & $\pm$&$ \uloggst$ & B \\
\meh & $\fe$ & $\pm$&$ \ufe$ & B \\
Distance~[pc] & \distance & $\pm$&$ \udistance$ & D \\
$T_{\rm eff}$ [K] & \teff & $\pm$&$ \uteff$ & B\\
 & & \\
\emph{Planet Parameters} & & & \\
$M_P$~[\mearth] & \mpl  &   $\pm$&$ \umpl$  & B,C,D \\
$R_P$~[\rearth] & \rpl &   $\pm$&$ \urpl$  & B,C,D \\
Mean planet density, $\rho_p$~[g cm$^{-3}$] & \rhopl & $\pm$&$ \urhopl$ & B,C,D \\
$\log g_p$~[cgs] & \loggpl  & $\pm$&$ \uloggpl$ & B,C,D \\
Equilibrium Temperature $T_{\rm eff}(\frac{R_\star}{2a})^{1/2}$~[K] & \teq & $\pm$&$\uteq$ &   B,C,D,E
\enddata

\tablecomments{(A) Determined from our analysis of the K2 light curve, the HARPS-N radial velocity measurements, the MOST light curve, and the stellar parameters. (B) Based on our spectroscopic analysis of the HARPS-N spectra. (C) Based on group A parameters.  (D) Based on the \Hipparcos\ parallax. (E) Assuming albedo of zero, and perfect heat redistribution.}

\end{deluxetable*}

We conducted an analysis of the K2 light curve, the HARPS-N radial velocity observations, and the MOST light curve and the WASP light curve to determine orbital and planetary properties. We first re-processed the K2 data using a different method from that described in VJ14 to minimize the possibility of any bias due to using the in--transit points in the flat field. We re-derived the SFF correction by fitting the K2 light curve using an MCMC algorithm with an affine invariant ensemble sampler \citep[adapted for IDL from the algorithm of ][]{goodman, emcee}. We fit the transit light curve with a \citet{mandelagol} model, as implemented by \citet{exofast}, with quadratic limb darkening coefficients held at the values given by \citet{claret}. We modeled the stellar out--of--transit variations with a cubic spline between 10 nodes spaced evenly in time, the heights of which were free parameters. Similarly, we modeled the SFF correction as a cubic spline with 15 nodes spaced evenly in ``arclength,'' a one--dimensional metric of position on the detector as defined in VJ14. Upon finding the best-fit parameters for the SFF correction, we applied the correction to the raw K2 data to obtain a debiased light curve.

After rederiving the correction to the K2 light curve, we simultaneously fit a transit light curve to the K2 light curve, the HARPS-N radial velocities, and the MOST and WASP photometry using {\em emcee}. We modeled the radial velocity variations with a two--planet Keplerian model (fitting the 9.1 day period and the 45 day period simultaneously), and modeled the transits of the 9.1 day planet with a \citet{mandelagol} model. For the K2 light curve, we accounted for the 29.4 minute long cadence exposure time by oversampling the model light curve by a factor of 13 and binning. We allowed limb darkening coefficients \citep[parametrized as suggested by][]{kippingld} to float. We used a white noise model for the radial velocity observations with a stellar jitter term added in quadrature to the HARPS-N formal measurement uncertainties. For the light curves, we used a Gaussian process noise model, using the same kernel described in Equation \ref{kernel}. We used an informative prior on the stellar $\log{g_\star}$ from Section \ref{stellarproperties} in our fits, which we converted to stellar density to help break the degeneracy between the scaled semimajor axis and impact parameter. Using this prior let us constrain the impact parameter despite having only one K2 long cadence datapoint during transit ingress and egress. In total, the model had 28 free parameters. We used \emcee\ to sample the likelihood space with 500 ``walkers'', each of which we evolved through 1500 steps. We recorded the last 300 of of these steps as samples of our posterior distribution, giving a total of 150000 MCMC links. We calculated correlation lengths for all 28 parameters which ranged from 5.6 to 19.0, corresponding to between 8000 and 27000 independent samples per parameter. We assessed the convergence of the MCMC chains using the spectral density diagnostic test of \citet{geweke} and found that the means of the two sequences are consistent for 21/28 parameters (75\%) at the 1--$\sigma$ level, for 27/28 (96\%) at the 2--$\sigma$ level, and 28/28 at the 3--$\sigma$ level. These fractions are consistent with draws from a normal distribution, which is the expected behavior for the MCMC chains having converged.

In Table \ref{tab:bigtable}, we report the best--fitting planet and orbit parameters and their uncertainties for \thisplanet\ by calculating the median link for each parameter and 68\% confidence intervals of all links for each parameter, respectively. We summarize the priors used in the fits and the full model outputs in Table \ref{tab:combinedtable}, including nuisance parameters like the noise model outputs. 

We find that our data are best described by the presence of a planet with $R_p$ = \rpl\ $\pm$ \urpl\ \mearth, $M_p$ = \mpl\ $\pm$ \umpl\ \mearth, in a \perpl\ day orbit. While we find some evidence for an outer planet in the radial velocity measurements, we cannot conclusively claim its existence based on the data presently at our disposal.

\section{Discussion}

\subsection{Composition of \thisplanet}

\begin{figure*}
\epsscale{1}
  \begin{center}
      \leavevmode
\plotone{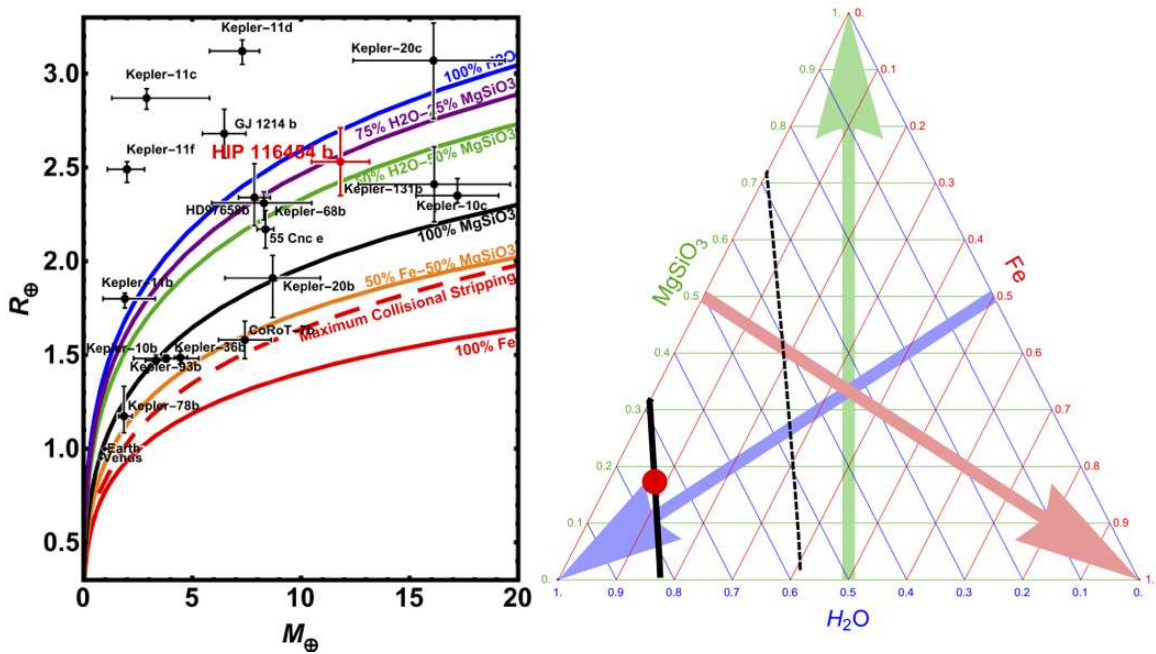}
\caption{Left: Mass/Radius diagram for sub-Neptune sized exoplanets. \thisplanet\ is consistent with being entirely solid, but has a density low enough that it could also have a substantial gaseous envelope surrounding a dense core. It is similar in mass, radius, and density to HD 97658b and Kepler 68b. Right: Ternary diagram showing allowed compositions for solid exoplanets. Assuming \thisplanet\ is solid, the thick dashed line represents allowed compositions for a planet with our best--fitting mass and radius, and the dashed line indicates the compositions allowed within one--$\sigma$ uncertainties. To one--$\sigma$, the planet must have at least 30\% water or other volatiles. } \label{massradius}
\end{center}
\end{figure*}

Figure \ref{massradius} shows \thisplanet\ on a mass-radius diagram with other known transiting, sub--Neptune--size exoplanets with measured masses and radii. We overlaid the plot with model composition contours from \citet{modelgrid}. We first note that \thisplanet\ has a mass and radius consistent with either a low--density solid planet, or a planet with a dense core and an extended gaseous envelope. The relatively low equilibrium temperature of the planet ($T_{eq} = \teq \pm \uteq$ K, assuming zero albedo and perfect heat redistribution) makes it unlikely that any gaseous envelope the planet started with would have evaporated over its lifetime. 

In terms of mass and radius, this planet is similar to Kepler-68\,b \citep[8.3 \mearth, 2.31 \rearth,][]{kep68} and HD\,97658\,b \citep[7.9 \mearth, 2.3 \rearth][]{howard, dragomir}, but is likely slightly larger. \thisplanet\ and HD\,97658\,b ($T_{\rm eq} \approx 730$ K) have similar effective temperatures while, Kepler-68\,b ($T_{\rm eq} \approx 1250$ K under the same assumptions) is somewhat hotter. Like these planets, \thisplanet\ has a density intermediate to that of rocky planets and that of ice giant planets. On the mass-radius diagram, \thisplanet\ lies close to the 75\% \water-25\% \mgsi\ curve for solid planets. It could be either a low--density solid planet with a large fraction of \water\ or other volatiles (which have similar equations of state to \water), or it could have a dense core with a thick gaseous layer.

We made inferences about the structure and composition of \thisplanet\ if it indeed has a dense core and thick gaseous envelope using analytic power law fits to the results of \citet{lopezfortney14}. Assuming the thick gaseous envelope is composed of hydrogen and helium in Solar abundances, an equilibrium temperature calculated with perfect heat redistribution and an albedo of 0, and a stellar age of about 2 Gyr, the models predict that \thisplanet\ has a hydrogen/helium envelope making up about 0.5\% of the planet's mass. The model suggests \thisplanet\ to have a 1.8 \rearth\ core with virtually all of the planet's mass, surrounded by a gaseous envelope with thickness 0.35 \rearth, and a radiative upper atmosphere also with thickness 0.35 \rearth. Using different assumptions to calculate the equilibrium temperature like imperfect heat distribution and a nonzero albedo \citep[for instance, the value of][]{sheets}, and different assumptions about the envelope's composition and age does not change the calculated thickness and mass of the gaseous envelope by more than a factor of two. We note that this envelope fraction is consistent with the population of \Kepler\ super-Earth/sub-Neptune sized planets studied by \citet{wolfgang}, who found the envelope fraction of these candidates to be distributed around 1\% with a scatter of 0.5 dex.

We also explored the composition of \thisplanet\ assuming it is solid and has little in the way of a gaseous envelope using an online tool\footnote{\url{http://www.astrozeng.com}}, based on the model grids of \citet{modelgrid}. We investigated a three--component model, with layers of \water, \mgsi\ and Fe. In this case, \thisplanet\ must have a significant fraction of either \water\ or other volatiles in an ice form like methane or ammonia. The composition in this case would be more similar to the ice giants in the Solar system than the rocky planets like Earth. 

We find that the pressure at the core of the planet can range from 1400~GPa for an iron--free planet to 2800~GPa for a silicate--free planet. Assuming a ratio of iron to silicates similar to that of the Earth and other Solar system bodies, we find that the core pressure of \thisplanet\ is about 2400~GPa. Under this assumption, \thisplanet\ would consist of 8\% Fe,17\% \mgsi\, and 75\% \water\ by mass. Using the ratio of iron to silicates in Solar system bodies is usually a relatively good assumption because this ratio is largely determined by element synthesis cosmochemistry, which does not vary greatly on the scale of 50 parsecs. However, \thisstar's unusual space motion indicates that it might have formed elsewhere in the galaxy, so this assumption might not hold. More detailed spectral analysis, in particular measuring elemental abundances for Mg and Si compared to Fe in the parent star could put additional constraints on the composition of \thisplanet\, assuming it is solid. 

If solid, \thisplanet\ would be one of the ``\water\ rich'' planets described in \citet{zengsasselov14}, for which it is possible to make inferences about the phase of the planet's \water\ layer, given knowledge of the star's age. Various evidence points to \thisstar\ having  an age of approximately 2 Gyr. Using relations from \citet{mamajek}, the $R'_{HK}$ level indicates an age of 2.7 Gyr and the rotation indicates an age of 1.1 Gyr, if the rotation period is indeed close to 16 days. The white dwarf's cooling age, however, sets a lower limit of approximately 1.3 Gyr. Future observations, like a mass measurement of the white dwarf to estimate its progenitor's mass (and therefore age on the main sequence), could constrain the age further. If \thisstar's age is indeed about 2 Gyr and the planet lacks a gaseous envelope, then it is likely to have water in plasma phases near its water-silicate boundary (the bottom of the \water\ layer), while if it is slightly older ($\sim$ 3 Gyr or more), or has a faster cooling rate, it could have superionic phases of water. 

\subsection{Suitability for Follow--up Observations}

\thisplanet\ is a promising transiting super-Earth for follow--up observations due to the brightness of its star, especially in the near infrared. We used the Exoplanet Orbit Database\footnote{\url{http://www.exoplanets.org}} \citep{exoplanets, exoplanets2} to compare \thisplanet\ to other transiting sub-Neptune sized planets orbiting bright stars. We found that among stars hosting transiting sub-Neptunes with $R_p < 3 R_\earth$, only Kepler 37, 55 Cnc, and HD 97658 have brighter $K$--band magnitudes.

\thisstar\ is particularly well--suited for additional follow--up photometric and radial velocity observations both to measure the mass of the planet to higher precision and to search for more planets in the system. \thisstar\ is chromospherically inactive and has low levels of stellar radial velocity jitter (0.45 $\pm$ 0.29 \ms). This combined with its brightness makes it an efficient radial velocity target. Moreover, the brightness of the host star makes \thisstar\ ideal for follow--up with the upcoming CHEOPS mission \citep{cheops}.

\thisplanet\ could be important in the era of the James Webb Space Telescope (JWST) to probe the transition between ice giants and rocky planets. In the Solar system, there are no planets with radii between 1--3~\rearth\, while population studies with \Kepler\ data have shown these planets to be nearly ubiquitous \citep{howardrates, fressin, petigura, mortonswift}. Atmospheric studies with transit transmission spectroscopy can help determine whether these planets are in fact solid or have a gaseous envelope, and give a better understanding on how these planets form and why they are so common in the Galaxy. Also of interest is the fact that \thisplanet\ is  very similar to HD\,97658\,b, in terms of its orbital characteristics (both are in $\sim$ 10 day low-eccentricity orbits), mass and radius (within 10\% in radius, and within 25\% in mass), and stellar hosts (both orbit K--dwarfs). Comparative studies of these two super--Earths will be valuable for understanding the diversity and possible origins of close--in Super--Earths around Sun--like stars. This being said, despite \thisstar's brightness, the relatively shallow transit depth will make it a somewhat less efficient target than super-Earths orbiting smaller stars \citep[for instance, GJ 1214 b, ][]{gj1214}. 
 
\subsection{Implications for K2 Science}

\thisplanet\ has demonstrated the potential of K2 to increase the number of bright transiting planets amenable to radial--velocity follow--up. Despite its degraded pointing precision, it is possible to calibrate and correct K2 data to the point where super-Earths can be detected to high significance with only one transit. Despite the increased expense of bright stars in terms of Kepler target pixels required for the aperture, K2 data is of high enough precision to produce many transiting exoplanets around bright stars. 

Many K2 fields, including the Engineering test field, are located such that observatories in both hemispheres can view the stars, a significant difference between K2 and the original \Kepler\ Mission. Even though all of our follow--up observations for \thisplanet\ took place at Northern observatories, the star's equatorial location enables follow--up from southern facilities like the original HARPS instrument at La Silla Observatory \citep{harps} and the Planet Finding Spectrograph at Las Campanas Observatory  \citep{pfs} just as easily as with Northern facilities with instruments like HARPS-N or the High Resolution Echelle Spectrograph at Keck Observatory \citep{hires}. 

Many \Kepler\ planet candidates were confirmed in part thanks to precise measurements of the \Kepler\ image centroid as the planet transited---the expected motion of the image centroid could be calculated based on the brightness and position of other stars near the aperture, and deviations from that prediction could signal the presence of a false--positive planet candidate. Such an analysis will be substantially more difficult for K2 data, because the unstable pointing leads to large movements of the image centroid. In this work, we were able to exclude the possibility of background objects creating false transit signals taking advantage of the star's high proper motion and archival imaging. This will be more difficult for more distant stars. However, the focus of the K2 mission on nearby late K-- and M--dwarfs, which typically have high proper motions, could make this technique of background star exclusion more widely applicable than it was for the original \Kepler\ mission.

\acknowledgments
We thank Ball Aerospace and the \Kepler/K2 team for their brilliant and tireless efforts to make the K2 mission a possibility and a success. Without their work, this result would not have been possible. We thank Sarah Ballard and Kevin Apps for helpful conversations. We acknowledge many helpful comments from our anonymous reviewer, as well as from Eric Feigelson, our scientific editor. 

Some of the data presented in this paper were obtained from the Mikulski Archive for Space Telescopes (MAST). STScI is operated by the Association of Universities for Research in Astronomy, Inc., under NASA contract NAS5--26555. Support for MAST for non--HST data is provided by the NASA Office of Space Science via grant NNX13AC07G and by other grants and contracts.This paper includes data collected by the \Kepler\ mission. Funding for the \Kepler\ mission is provided by the NASA Science Mission directorate.

This research has made use of NASA's Astrophysics Data System, the SIMBAD database and VizieR catalog access tool, operated at CDS, Strasbourg, France, the Exoplanet Orbit Database and the Exoplanet Data Explorer at \url{http://www.exoplanets.org}, PyAstronomy, the repository and documentation for which can be found at \url{https://github.com/sczesla/PyAstronomy}, and the NASA Exoplanet Archive, which is operated by the California Institute of Technology, under contract with the National Aeronautics and Space Administration under the Exoplanet Exploration Program.

A.V. and B.T.M are supported by the National Science Foundation Graduate Research Fellowship, Grants No. DGE 1144152 and DGE 1144469, respectively. J.A.J is supported by generous grants from the David and Lucile Packard and Alfred P. Sloan Foundations. C.B. acknowledges support from the Alfred P. Sloan Foundation. PF acknowledges support by  Funda\c{c}\~ao para a Ci\^encia e a Tecnologia (FCT) through Investigador FCT contracts of reference IF/01037/2013 and POPH/FSE (EC) by FEDER funding through the program ``Programa Operacional de Factores de Competitividade - COMPETE''. WWW was supported by the Austrian Science Fund (FWF P22691-N16). The research leading to these results has received funding from the European Union Seventh Framework Programme (FP7/2007-2013) under Grant Agreement n. 313014 (ETAEARTH). This publication was made possible through the support of a grant from the John Templeton Foundation. The opinions expressed in this publication are those of the authors and do not necessarily reflect the views of the John Templeton Foundation.

This work is based on observations made with the Italian Telescopio Nazionale Galileo (TNG) operated on the island of La Palma by the Fundación Galileo Galilei of the INAF (Istituto Nazionale di Astrofisica) at the Spanish Observatorio del Roque de los Muchachos of the Instituto de Astrofisica de Canarias. The HARPS-N project was funded by the Prodex Program of the Swiss Space Office (SSO), the Harvard University Origin of Life Initiative (HUOLI), the Scottish Universities Physics Alliance (SUPA), the University of Geneva, the Smithsonian Astrophysical Observatory (SAO), and the Italian National Astrophysical Institute (INAF), University of St. Andrews, Queens University Belfast and University of Edinburgh.

The Robo-AO system is supported by collaborating partner institutions, the California Institute of Technology and the Inter-University Centre for Astronomy and Astrophysics, and by the National Science Foundation under Grant Nos. AST-0906060, AST-0960343, and AST-1207891, by the Mount Cuba Astronomical Foundation, by a gift from Samuel Oschin.

Some of the data presented herein were obtained at the W.M. Keck Observatory, which is operated as a scientific partnership among the California Institute of Technology, the University of California and the National Aeronautics and Space Administration. The Observatory was made possible by the generous financial support of the W.M. Keck Foundation. The authors wish to recognize and acknowledge the very significant cultural role and reverence that the summit of Maunakea has always had within the indigenous Hawaiian community.  We are most fortunate to have the opportunity to conduct observations from this mountain. 

WASP-South is hosted by the SAAO and SuperWASP by the Isaac Newton Group and the Instituto de Astrof\'isica de Canarias; we gratefully acknowledge their ongoing support and assistance. Funding for WASP comes from consortium universities and from the UK’s Science and Technology Facilities Council (STFC).

The Digitized Sky Surveys were produced at the Space Telescope Science Institute under U.S. Government grant NAG W-2166. The images of these surveys are based on photographic data obtained using the Oschin Schmidt Telescope on Palomar Mountain and the U.K. Schmidt Telescope. The plates were processed into the present compressed digital form with the permission of these institutions.

The National Geographic Society--Palomar Observatory Sky Atlas (POSS-I) was made by the California Institute of Technology with grants from the National Geographic Society. The Second Palomar Observatory Sky Survey (POSS-II) was made by the California Institute of Technology with funds from the National Science Foundation, the National Geographic Society, the Sloan Foundation, the Samuel Oschin Foundation, and the Eastman Kodak Corporation. The Oschin Schmidt Telescope is operated by the California Institute of Technology and Palomar Observatory.

Funding for SDSS-III has been provided by the Alfred P. Sloan Foundation, the Participating Institutions, the National Science Foundation, and the U.S. Department of Energy Office of Science. The SDSS-III web site is http://www.sdss3.org/.

SDSS-III is managed by the Astrophysical Research Consortium for the Participating Institutions of the SDSS-III Collaboration including the University of Arizona, the Brazilian Participation Group, Brookhaven National Laboratory, Carnegie Mellon University, University of Florida, the French Participation Group, the German Participation Group, Harvard University, the Instituto de Astrofisica de Canarias, the Michigan State/Notre Dame/JINA Participation Group, Johns Hopkins University, Lawrence Berkeley National Laboratory, Max Planck Institute for Astrophysics, Max Planck Institute for Extraterrestrial Physics, New Mexico State University, New York University, Ohio State University, Pennsylvania State University, University of Portsmouth, Princeton University, the Spanish Participation Group, University of Tokyo, University of Utah, Vanderbilt University, University of Virginia, University of Washington, and Yale University.

Facilities: \facility{Kepler, MOST, FLWO:1.5m (CfA Digital Speedometer, TRES), TNG (HARPS-N), PO:1.5m (Robo-AO), PO:1.2m, Keck:II (NIRC2)}


\clearpage

\begin{deluxetable*}{lcccc}
\tablecaption{Summary of Combined Analysis \label{tab:combinedtable}}
\tablewidth{0pt}
\tablehead{
  \colhead{Parameter} & 
  \colhead{Prior}     &
  \colhead{50\% Value} &
  \colhead{15.8\% }     &
  \colhead{84.2\% }   
}
\startdata
$\sqrt{e_1}\cos(\omega_1)$ & $\mathcal{U}(-1, 1)$ & 0.244 & -0.052 &  +0.049 \\ 
$\sqrt{e_1}\sin(\omega_1)$ & $\mathcal{U}(-1, 1)$ & -0.395 & -0.091 &  +0.109 \\ 
$t_{t,1}$, [BJD] & $\mathcal{U}(-\infty, \infty)$ & 2456907.895 & -0.037 &  +0.015 \\ 
$\log{(P_1/day)}$ & $\mathcal{U}(-\infty, \infty)$ & 2.210472 & -0.000176 &  +0.000079 \\ 
$\log{(M_1/M_{\rm Jup})}$ & $\mathcal{U}(-\infty, \infty)$ & -3.286 & -0.111 &  +0.089 \\ 
$\cos{i_1}$ & $\mathcal{U}(-1, 1)$ & -0.0296 & -0.0038 &  +0.0023 \\ 
$\sqrt{e_2}\cos(\omega_2)$ & $\mathcal{U}(-1, 1)$ & 0.19 & -0.26 &  +0.22 \\ 
$\sqrt{e_2}\sin(\omega_2)$ & $\mathcal{U}(-1, 1)$ & -0.581 & -0.122 &  +0.186 \\ 
$t_{t,2}$, [BJD] & $\mathcal{U}(-\infty, \infty)$ & 2456930.8 & -5.4 &  +5.7 \\ 
$\log{(P_2/day)}$ & $\mathcal{U}(-\infty, \infty)$ & 3.838 & -0.082 &  +0.093 \\ 
$\log{(M_2 \sin{i_2}/M_{\rm Jup})}$ & $\mathcal{U}(-\infty, \infty)$ & -2.22 & -0.29 &  +0.28 \\ 
$\log{R_{p,1}/R_\star}$ & $\mathcal{U}(-\infty, \infty)$ & -3.443 & -0.039 &  +0.042 \\ 
$q_1$ & $\mathcal{U}(0, 1)$ & 0.37 & -0.24 &  +0.33 \\ 
$q_2$ & $\mathcal{U}(0, 1-q_1)$ & 0.41 & -0.39 &  +0.29 \\ 
RV Zero Point [ms$^-1$] & $\mathcal{U}(-\infty, \infty)$ & -0.42 & -0.34 &  +0.33 \\ 
Jitter [ms$^-1$] & $\mathcal{U}(-\infty, \infty)$ & 0.45 & -0.28 &  +0.32 \\ 
$\log{g_\star}$ & $\mathcal{N}(4.59, 0.026)$ & 4.591 & -0.020 &  +0.022 \\ 
$\log{A}$ & $\mathcal{U}(-20, 5.5)$ & -13.05 & -0.33 &  +0.35 \\ 
$\log{l}$ & $\mathcal{U}(-2, 8.5)$ & 1.33 & -0.93 &  +1.33 \\ 
$\log{g_{q}}$ & $\mathcal{U}(-8, 7)$ & 2.7 & -1.6 &  +1.4 \\ 
$\log{(P_{\rm rot}/day)}$ & $\mathcal{U}(2, 4)$ & 3.31 & -0.78 &  +0.53 \\ 
$\log{\sigma_{\rm WASP1}}$ & $\mathcal{U}(-6.75, -1.75)$ & -4.1 & -1.8 &  +1.8 \\ 
$\log{\sigma_{\rm WASP2}}$ & $\mathcal{U}(-6.75, -1.75)$ & -4.2 & -1.8 &  +1.8 \\ 
$\log{\sigma_{\rm WASP3}}$ & $\mathcal{U}(-6.75, -1.75)$ & -4.3 & -1.7 &  +1.9 \\ 
$\log{\sigma_{\rm K2}}$ & $\mathcal{U}(-11, -6)$ & -9.284 & -0.044 &  +0.048 \\ 
$\log{\sigma_{\rm MOST1}}$ & $\mathcal{U}(-9, -3)$ & -6.685 & -0.059 &  +0.064 \\ 
$\log{\sigma_{\rm MOST2}}$ & $\mathcal{U}(-9, -3)$ & -7.227 & -0.051 &  +0.051 \\ 
$\log{\sigma_{\rm MOST3}}$ & $\mathcal{U}(-9, -3)$ & -6.710 & -0.122 &  +0.149 \\
\enddata

\tablecomments{$\mathcal{U}(A, B)$ represents a uniform distribution between A and B, and $\mathcal{N}(\mu, \sigma)$ represents a normal distribution with mean $\mu$ and standard deviation $\sigma$. The limb darkening coefficients $q_1$ and $q_2$ are defined according to the parameterization of \citet{kippingld}. All logarithms are base $e$.}

\end{deluxetable*}

\end{document}